\documentclass[preprint,12pt]{elsarticle}

\usepackage{graphicx}
\usepackage{amsmath}
\usepackage{amssymb}
\usepackage{amsthm}
\usepackage[slantedGreek]{mathptmx}
\usepackage{url}
\usepackage{bbm}
\usepackage{accents}

\newtheorem{lemma}{Lemma}

\newtheorem*{theorem*}{Theorem}

\journal{Theoretical Economics}

\begin{document}

\begin{frontmatter}

\title{RPS(1) preferences}
\author{Misha Perepelitsa }

\date{\today}
\address{
misha@math.uh.edu\\
Department of Mathematics\\
University of Houston\\
4800 Calhoun Rd. \\
Houston, TX.}

\begin{abstract}
We consider a model for decision making  based on 
an adaptive, $k$--period,  learning process where  the priors are selected according to  Von Neumann-Morgenstern expected utility principle. A preference relation between two prospects is introduced, depending on which prospect is selected more often.  We show that the new preferences have similarities with the preferences  obtained by Kahneman and Tversky (1979) in the context of the prospect theory. Additionally,  we establish that in the limit of large learning period $k,$ the new preferences  coincide with  the expected utility principle.

\end{abstract}


\end{frontmatter}

\begin{section}{Introduction}

The expected utility  theory (EUT) was put forward by Von Neumann and Morgenstern (1947) as a mathematical formalization of the way the choices are made between the prospects involving objective, probabilistic outcomes. The theory derives a utility function $U$ (VNM utility)  which  assigns values to the payoffs and  is used to rank prospects according to the expectation $\mathbb{E}[U(X)],$ where $X$ is a random variable of payoffs of  a given prospect.

 As a  theory, EUT  starts with the set of axioms, called axioms of choice that one has to adhere in assigning the preferences to prospects. They are the  completeness, transitivity, continuity and independence (substitution) axioms. The last axioms drew a significant amount of critique from the experimentalists, starting with Allais (1953).  Over the years several alternative utility theories were proposed  that provide some variants of the expected utility without the independence axiom or with its weaker  version. Among them, the generalized expected utility of Machina (1982), weighted utility theory of Chew and MacCrimmon (1979),
the regret theory developed independently by   Bell (1982), Fishburn (1982),  Loomis and Sugden (1982), rank dependent utility theory of Quiggin (1982, 1993), and the dual utility theory by Yaari (1987).
 
Kahneman and Tversky (1979, 1984, 1992) introduced  framing effects, value functions, and  probability weights into the analysis  and incorporated them into the prospect theory that was later developed, using the 
approach of the rank dependent utility theory, into the cumulative prospect theory.

In this paper we will consider the deviations from the EU principle due to                                                                                                                                                                                                                                                      response to the current incentives, as modeled by a reinforcement learning mechanism.  Rather than postulating axioms of choice we will postulate the type of behavior that directs an agent to a particular choice, and  derive the preferences for prospects from that. What we have in mind here is a reactive response to outcomes, as seen, for example,  when people are rushing to buy  flood insurance just after a historic flood, and then, canceling it in a few years. To an objectivist, who believes in statistical statements like ``1 in 100 years flood plan,'' this behavior is ``irrational.'' However if a degree of uncertainty about the future is allowed (weather pattern is changing) the better name for it is ``adaptive,'' which, when coupled with a learning process that accounts for accumulating experience, represents a ubiquitous behavioral pattern. 

The relevance of such behaviors for the problems in Economics has always been acknowledged. A systematic approach has been developed by Cross  in the monograph ``A theory for adaptive economic behavior'' (1983). In that work, the author analyzes  stochastic models for the decision making built on the principles of the learning theory laid out by Bush and Mosteller (1951, 1955).  Agents actions evolve through a feedback mechanism that  updates the  choice probabilities. The focus is on the properties of  choice and behavior in the transient regime (called disequilibrium), starting at the moment when the exogenous parameters have changed, and prior to the moment the rational equilibrium is reached in the long run.

In this paper we will extend the Cross' theory to describe the preferences  for risky prospects that can be associated with adaptive behavior. This is done by changing the feedback model of Cross to a short-term memory model, when the actions are determined only from few previous experiences. The approach allows us to derive exact expressions for preferences and compare them with the preferences based on the Expected Utility principle.

The new type of preferences will be build upon an agent's VNM utility function $U$ and the expected utilities of prospects, calculated according to EUT.  We will assume that the agent does {\it rank} prospects according to EU, but  the agent's {\it actions} are not determined solely by the expected utilities. They rather depend on the combination of factors expressed by the 
  level of {\it stimulus} (propensity) for selecting a prospect. The stimulus for a prospect  will be defined as  a sum of the ``default level of stimulus'' (stimulus prior), expressed by its expected utility and the increments, denominated in the units of the utility function, from the most recent payoffs the prospect paid out.  At this point, it is best to think  of a dynamic (iterated) decision making process, in which the agent constantly updates her attitude toward the prospects.  Given the stimuli for prospects, the agent  chooses the prospect with a probability, defined as a fraction of
the prospect stimuli  to the total stimuli of all prospect.  The new preference relation between two prospects $L_a$ and $L_b$ is defined by a ranking based on {\it which prospect is selected more often in a long run of choices.} We will call such relation RPS (relative payoff sum) preferences, in analogy the learning algorithm, introduced by Harley (1981) in the context of the Game Theory. In that filed, the relation between the reinforcement learning and optimal strategies was studied by  Roth and Erev (1995, 1998), B\"orgers and Sarin (1995, 1997), Erev and Roth (1996), and Fudenberg and Levine (1998).


The new preferences are derived by including adaptation mechanism into decision making. In the  view of this, it is particular interesting to see that  RPS preferences follow the pattern of violations to the EU theory observed experimentally.  Namely, RPS preferences are non-linear, in way similar to the preferences in  Allais' paradox. They are frame dependent, and the risk attitude, in comparison to the  EU risk attitude, changes from  more risky to more conservative, as payoffs shift from losses to gains. 

\end{section}

\begin{section}{RPS models of decision making}

To motivate the new type of the preferences we consider the decision making as a dynamics ``learning process''. Given the choice between two prospects (lotteries) $L_a$ and $L_b,$  an agent uses her VNM utility function to compute the expected utilities  $U_a$ and $U_b$ to  compare the prospects. Suppose now that the agent has to select a prospect on a regular basis. Even though the agent has a preference for one, she might be influenced by a variety of other factors (mood, peer-pressure, fashion trends, etc.) to select less preferred prospect at a given instant of time. Those factors are not included in the evaluation of the utility, because, typically, they are too complex to account for  and/or unpredictable.

A customary approach to model this situation is to assume that the agent follows a mixed strategy, by selecting one prospect or another with a certain probability, which, in the view of the given conditions, can be set to be the proportion of utilities, i.e. the probability to select prospect $L_a$ is determined by the ratio
\begin{equation}
\label{prob:U}
\frac{U_a}{U_a+U_b}.
\end{equation}
In this way both prospects are being repeatedly selected. The important property of such dynamic decision process is its consistency with the EU principle:  the prospect that is selected most often is the prospect with the higher expected utility.

To model the deviations from the EU principle, we will assume that the agent adjusts the probability \eqref{prob:U} from one decision to another depending on the history of  realized payoffs. The adjustment is modeled by a positive/negative reinforcement mechanism.

First, we consider  the reinforcement model  determined only by the last payoff.
\subsection{RPS(1) preferences}
To simplify the presentation, we consider lotteries with  non-negative payoffs first, and relax this condition in subsequent sections.  

Let $P_a^i,\,P_b^i$ be the payoffs from lotteries $L_a$ and $L_b$ at epoch $i.$ The reinforcement mechanism is defined
through the stimuli (propensities) to select $L_a$ and $L_b$ at the next epoch $i+1.$ They are set to be proportional to the utilities of the last payoff: 
\begin{eqnarray}
\label{s.1}
S_a^{i+1}{}={}U_a + U(P_a^{i}), \\
\label{s.2}
S_b^{i+1}{}={}U_b + U(P_b^{i}),
\end{eqnarray}
where,  we set expression $U(P_a^{i}),$ or $U(P_b^{i}),$ to zero if the lottery is not selected. $U_a,$ $U_b$ are the default levels of the stimuli.  At the next epoch the agent executes a mixed strategy by selecting  lottery $L_a$ with the decision probability 
\begin{equation}
\label{prob}
\lambda_a^{i+1}=\frac{S_a^{i+1}}{S_a^{i+1}+S_b^{i+1}}. 
\end{equation}
Once the lottery is selected, the payoff is determined by a random sampling of lottery $L_a$ or $L_b.$

We will call rule \eqref{s.1}--\eqref{prob} the relative payoff sum (or RPS(1))  decision making model, following the nomenclature introduced by Harley (1981). Note that this model is a {\it short memory} model, unlike most of the learning models considered in the literature cited in the Introduction. 


In the course of repeated choices the agent will keep switching between two lotteries. A statistically minded agent however, will notice that one lottery is being selected more often than the other. With more choices made, this observation gets re-confirmed.

It leads to  a preference relation, $L_a\succ L_b,$  defined by the rule  that {\it lottery $L_a$ is selected, in the long run,  more often than $L_b,$ when the decisions are made according to the rule \eqref{s.1}--\eqref{prob}.}   

As shown in Appendix, the mathematical equivalent of this definition is inequality
\begin{equation}
\label{RPS}
\mathbb{E}\left[\frac{U_b}{U_a+U_b+U(X)}\right] {}<{}\mathbb{E}\left[\frac{U_a}{U_a+U_b+U(Y)}\right],
\end{equation}
where $X$ is a random payoff corresponding to lottery $L_a$ and $Y$ is that of $L_b.$ We call this relation RPS(1) preferences. 
These preferences can also be interpreted as ``expected decision probability'' principle,  according to formula \eqref{A:RPS} from Appendix. 

In terms of the agent's cognition, new preferences correspond to the following mental math calculations: ``To select between $L_a$ and $L_b,$ I will compare my average willingness to repeat the same choice next time if I selected each lottery now and observed its payoff.''

Here, by ``willingness'' we understand the decision probability \eqref{prob} calculated from the ratio of stimuli. Clearly, this rule is not restricted to the iterated decision making scenario described above, but can be  deliberated even when one-time decisions are made.

RPS(1) preferences differ form the corresponding EU preferences. This, and other properties, are discussed below, but a quick example can help to illustrate it. Consider a lottery that pays out \$100 for sure and a lottery
that pays out \$0 and \$200 with 50/50 chances. If the agent is risk neutral, with the linear utility $U(x)=x,$ then she is indifferent between two lotteries in EU preferences, but according to RPS(1)-rule \eqref{RPS}, the sure bet is more attractive to her.


 We list below  the properties of RPS(1) preferences that follow directly from the definition, with explanations provided in Appendix.
\begin{enumerate}
\item[a.] RPS(1) preferences are not invariant under the shifts of the utility function from $U(x)$ to $U(x)+a.$ They exhibit the  framing effect. This property is considered in more details in section \ref{losses}.

\item[b.] Given the priors $U_a$ and $U_b,$ RPS(1) preferences depend on wealth increments, not the total accumulated wealth.

\item[c.] RPS(1) preferences are not transitive.

\item[d.] RPS(1) preferences violate the independence axiom.

\item[e.] RPS(1) preferences verify first order stochastic dominance (FSD) principle. 

\end{enumerate}

\subsection{Independence Axiom and Allais experiment}
We illustrate the failure of the independence axiom on the following numerical example.
Consider a concave VNM utility $U(c)= (1+c)^{.2}-1,$ normalized so that $U(0)=0.$  Denote by $\hat{L}_c$ the  lottery that pays \$c for sure, and by $L_x$ the one that  pays \$0 with probability $1-x$ and \$1 with probability $x.$ We compare lotteries $L_c$ and $L_x$ by RPS(1)-rule. 
The result is represented graphically in figure \ref{fig:1}. The dashed line divides the unit square into two parts: below the line, the certain bet $\hat{L}_c\succ L_x,$ above, $L_x\succ \hat{L}_c,$ and the line itself is the indifference curve (certainty equivalent curve). 

Similar to Allais experiment, we mix each lottery, $L_x$ and $\hat{L}_c,$ with  80\% chance of lottery $L_0$ which pays \$0 for sure. In violation of the independence axiom, the preference change: the solid line in figure \ref{fig:1} is the new indifference curve. In the region between the two lines $\hat{L}_c \succ L_x,$ but   $0.2L_x+0.8L_0 \succ 0.2\hat{L}_c+0.8L_0.$ 

\begin{figure}[t]
\centering
\includegraphics[width=12cm]{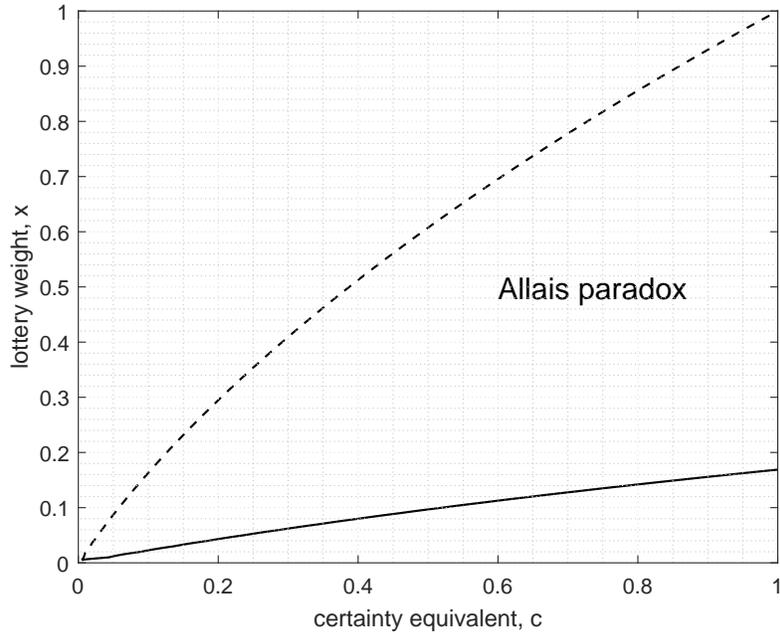}
\caption{Allais Paradox. The dashed line is the certainty equivalent for lottery $L_x$  in RPS(1) preferences, when  VNM utility $U(c)=(1-c)^{0.2}-1.$  The solid line represents the points $(c,x)$ for which lottery $0.2L_x+0.8L_0 \sim 0.2\hat{L}_c+0.8L_0.$ The region between the curves is set of points for which Allais paradox holds. \label{fig:1}}

\end{figure}

\subsection{Losses and Gains}
\label{losses}

Consider now the situation when the wealth increments are framed as losses (negative) or gains (positive). The decision rule \eqref{prob} must be re-defined for expression \eqref{prob} to be meaningful. An approach adopted in learning literature cited in the Introduction is to  use a positive, non-decreasing function, for example, $\Phi(x)=e^{x/d}$, $d>0,$ and set
\[
\lambda^{i+1}_a{}={}\frac{\Phi(S^{i+1}_a)}{\Phi(S^{i+1}_a)+\Phi(S^{i+1}_b).}
\]
The preferences, that we still call RPS(1) preferences, are defined by inequality
\begin{equation}
\label{RPS-1}
\mathbb{E}\left[\frac{\Phi(U_b)}{\Phi(U_b)+\Phi(U_a+U(X))}\right] {}<{}\mathbb{E}\left[\frac{\Phi(U_a)}{\Phi(U_a)+\Phi(U_b+U(Y))}\right],
\end{equation} 
when $L_a \succ L_b.$ 

In this setting we are going to look at the problem of determining the certainty equivalent for probabilistic lotteries involving positive or negative payoffs. 

As an example, we take logarithmic utility $U(c) = \log(1+0.4c)/\log1.4,$ normalized so that $U(0)=0,$ $U(1)=1.$ First we consider lottery $L_x$ that pays \$0 with probability $1-x$ and \$1 with probability $x.$  For each such lottery we find its certainty equivalent \$c according to RPS(1)-preferences. Figure \ref{fig:2} shows, by the solid line, the corresponding indifference curve, in the first quadrant.  Then, we consider lotteries $L_x$ that pay \$0 with probability $1-x$ and $U(-1)<0$ with probability $x.$ The solid line in the third quadrant is the indifference curve for losses.  For comparison we draw on the same figure the certainty equivalent curve (dashed) according to the EU principle, which is simply the graph of $U(c).$

As seen from that figure, the indifference curve for RPS(1)-preferences  lies above $U(c)$ in the region of gains, and below it, for losses.  Thus, an agent is more risk averse, compared to the expected utility preferences, in gains, and more risk tolerant  in losses. 
It should be noted that the situation in figure \ref{fig:2} applies only to two-valued lotteries described there. Since RPS(1) is not an expected utility no conclusions about other types of lotteries can be drawn from the graph of the certainty equivalents of such lotteries. In particular, the type of convexity of the certainty equivalent curve can not be used to characterize RPS(1) preferences in the relation to the attitude toward risk.

However, as shown in Appendix, the following property is generic.
\begin{lemma} Suppose that  VNM utility is risk-neutral, $U(x)=x,$ and let $L_a$ be any lottery,
described by a payoff random variable $X.$  Then, if $X\geq 0,$ the RPS(1) certainty equivalent of $L_a$ is less than the expected utility $U_a=\mathbb{E}[U(X)].$ If $X\leq0,$
 the RPS(1) certainty equivalent of $L_a$ is greater than the expected utility $U_a.$
\end{lemma}



\begin{figure}[t]
\centering
\includegraphics[width=13cm]{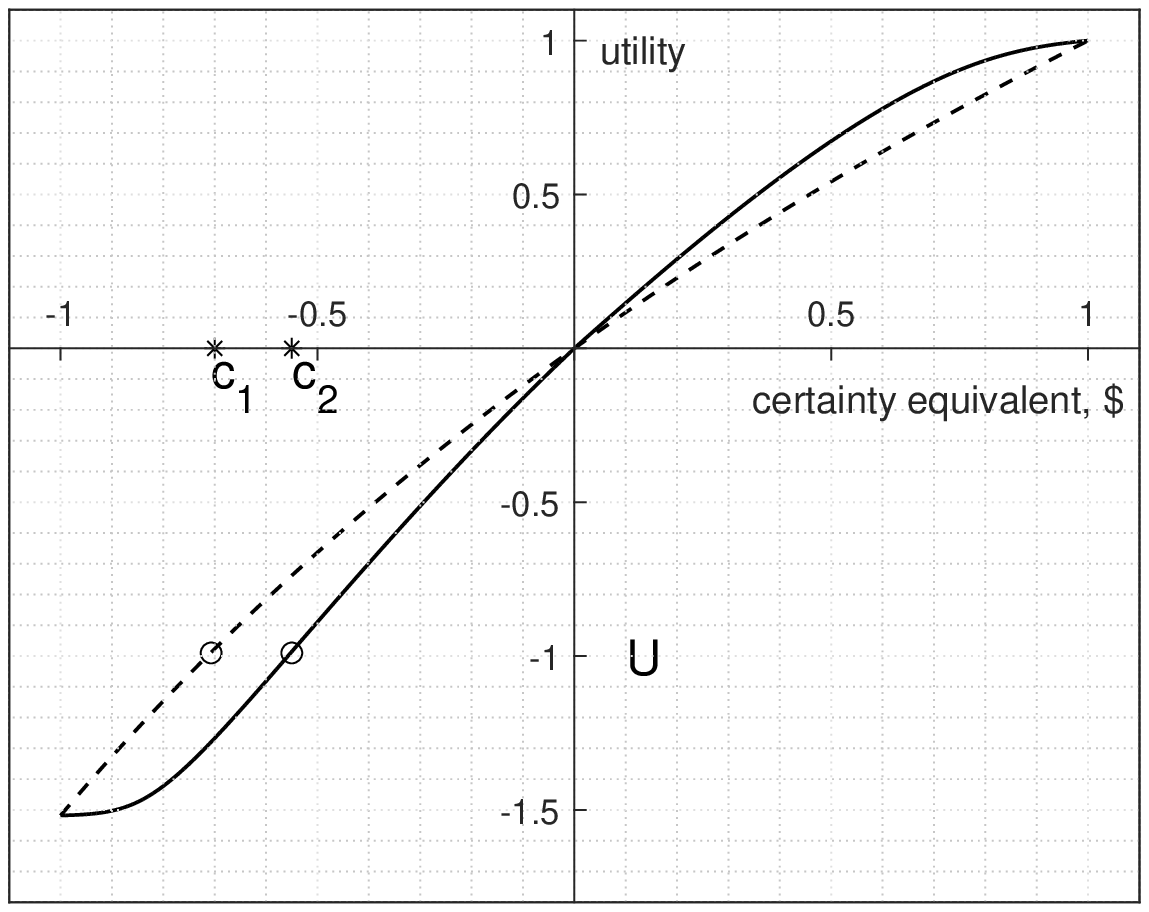}
\caption{Certainty equivalent curves for simple lotteries $L_x$ for losses and gains according to  EU (dashed) and RPS(1) (solid) preferences.  Utility $U(c) = \log(1+.4c)/\log1.4.$ Given the level of utility $U>0,$ lottery $L_x$  pays  either \$0 or \$1, so that $U=(1-x)U(0)+xU(1).$ If $U<0,$ $L_x$  pays either \$0 or \$-1, so that $U=(1-x)U(0)+xU(-1),$ with the corresponding certainty equivalents: $\$c_1$ is certainty equivalent in EU preferences, and $\$c_2$ in RPS(1).  For EU preferences (dashed line) the curve is
the graph of $U(c).$ The function that defines the stimuli, $\Phi(x)=e^{x/.2}.$ \label{fig:2}}

It should be mentioned the a similar asymmetry between losses and gains in the adaptive behavior is observed in the original learning model of Cross (1983).

\end{figure}

\subsection{RPS(k)-preferences}
\label{sec:rps}
A direct extension of the decision model \eqref{s.1}--\eqref{prob} is a model in which an agent accounts for  the last $k$ payoffs. The stimuli change, for example, by  the average:
\begin{eqnarray}
\label{s.3}
S_a^{i+1}{}={}U_a + \frac{1}{n_a^i}\sum_{j=0}^{k-1}U(P_a^{i-j}), \\
\label{s.4}
S_b^{i+1}{}={}U_b + \frac{1}{n_b^i}\sum_{j=0}^{k-1}U(P_b^{i-j}),
\end{eqnarray}
where $n^i_a$ is the number of times lottery $L_a$ has been selected during the last $k$ rounds, and similarly for $n^i_b.$ Again, we are using the convention that if $L_a$ ($L_b$)
  was not chosen at round $i,$ then $U(P_a^i)=0$ ($U(P_b^i)=0$). The decision between lotteries is made randomly with decision probability \eqref{prob}.

We define the preference relation  $L_a \succ L_b$ if  $L_a$ is selected {\it more often} in a long series of decision making, and refer to such preferences as RPS(k)-rule. 

It is possible to develop a mathematical formalism similar to \eqref{RPS}, however it is outside of the scope of this work. Here, we would like to use RPS(k)-preferences to a establish a connections between the adaptive decision making and the expected utility theory.

Consider RPS(k)-preferences with the memory extending to utmost past, that is, $k\to+\infty.$  In the limit, the average payoffs in \eqref{s.3}, \eqref{s.4} become the expected utilities $U_a$ and $U_b,$ respectively. In this case, the lottery most often played is the one with the highest expected utility. Thus, the adaptive decision making in the long memory limit takes agents' adaptive RPS-preferences to the expected utility preferences.

The relation of the EU preferences and long-run learning is by no means original. It has been expressed by many authors, including Cross (1983) and Plott (1996).



\subsection{Discussion}
We live in changing environments. 
Some environments are fairly stationary and predictable, like changes of seasons. In the course of the human history we developed certain rules that regulate our decisions about such processes, like the decision to sow in spring and reap in fall, with little  variations.  Even the occurrence of  an unprecedentedly warm winter followed by a cold summer won't moves us  to act contrary to the rule and we won't invest our effort in planting seeds in the fall.  

Other processes, like the changes in the financial markets,  have a great deal of uncertainty in future outcomes.   If 
a large drop in a stock price is  observed today, should it be considered as a random fluke, an outlier of a probabilistic model one uses, or is it an indication that it is time to adjust the means, deviations and other parameters in the model, or even to sell all the stocks immediately? Most likely, if we lost a significant sum of money due to this drop, our first reaction will be to sell to avoid further losses, and we have to analyze more information and  do some non-trivial mental work to find arguments for not doing so, or to confirm our initial impulse.

Reaction to the current stimuli, as the one just described, is a part of our biological adaptive response. It has been developed in the course of the Evolution to let us quickly react once the environment has changed, increasing our chances for survival. 
To a higher degree, it manifests itself during  the critical events, such as  market crashes or social upheavals, 
when old rules of behavior become irrelevant and people pursue minute gains, resulting in a state of  chaos,  until new rules are learned or enforced.

It is not unlikely that this response to current stimuli is 
present in all decisions that we make, but it is balanced against the past experience. We suggest that it is  one of the main contributors to the deviations of the experimentally observed behaviors from EU principle. In this paper we considered a  simple decision making  model for such deviations, called RPS preferences. Moreover, we showed that the EU preferences can be derived from RPS preferences in infinitely long-memory adaptive behavior.

The point of view suggested here is  that decision making is {\it a dynamics learning process.} When placed in stationary environments, long-memory will lead to the formation of the preferences (rules) that  resemble EU preferences. The more advantageous these rules are, the more pronounced they affect the decisions that me make. This seems to be the conventional notion of rationality. However, ``the irrational part'', the response to current stimuli, is also always  present.  It becomes dominating when the environment is changing, and perhaps, is our best choice when the changes are completely unpredictable.

\end{section}
\begin{section}{Appendix}
\subsection{Equation \eqref{RPS}}
Consider the stochastic decision making process from section \ref{sec:rps}. Let $X^i$ denote the lottery, either $L_a,$ or $L_b,$ selected by the agent at step $i.$ It is a Markov chain on the state of two elements $\{L_a,\,L_b\}.$ Denote by $f(x)$ the probability density for the distribution of payoffs in lottery $L_a,$ and by $g(x)$ the probability density for $L_b.$ The matrix of transition probabilities for the Markov chain equals to  
\[
T{}={}\left[
\begin{array}{rr}
\displaystyle\int \frac{U_a+U(x)}{U_a+U_b+U(x)}f(x)\,dx & 1 - \displaystyle\int \frac{U_b+U(x)}{U_a+U_b+U(x)}g(x)\,dx \\[2ex]
1- \displaystyle\int \frac{U_a+U(x)}{U_a+U_b+U(x)}f(x)\,dx & \displaystyle\int \frac{U_b+U(x)}{U_a+U_b+U(x)}g(x)\,dx
\end{array}
\right]
\]

The Markov chain is irreducible and both states are ergodic, see the monograph of Feller (1957).  This implies that the distribution of $X^i$ converges to the invariant measure, given by the pair of probabilities $(p,q)$ solving the matrix equation
\[
\left[\begin{array}{r}p\\q\end{array}\right]{}={}T\left[\begin{array}{r}p\\q\end{array}\right].
\]
Lottery $L_a\succ L_b$ iff $p>q.$ From the above equation we find that it is equivalent to
\begin{equation}
\label{A:RPS}
\int \frac{U_a+U(x)}{U_a+U_b+U(x)}f(x)\,dx > \int \frac{U_b+U(x)}{U_a+U_b+U(x)}g(x)\,dx,
\end{equation}
which is equivalent to  \eqref{RPS}. In the view of the definition of the probability $\lambda$ in \eqref{prob}, the above inequality says that the expected probability to select $L_a,$ provided that $L_a$ was selected last time, is greater than the expected probability to select $L_b,$ provided that $L_b$ was selected last time.

\subsection{First order stochastic dominance}
Suppose that lottery $L_a,$ described by random payoff variable $X,$ stochastically dominates $L_b$ described by random payoff  $Y.$ Let $U(c)$ be a non-decreasing function. Then, 
\[
U_b{}={}\mathbb{E}[U(Y)] < U_a{}={}\mathbb{E}[U(X)],
\]
and 
\[
\mathbb{E}\left[ \frac{1}{U_a + U_b +U(X)}\right] < \mathbb{E}\left[ \frac{1}{U_a + U_b +U(Y)}\right].
\]
From these two inequalities it follows that 
\[
\mathbb{E}\left[ \frac{U_b}{U_a + U_b +U(X)}\right] < \mathbb{E}\left[ \frac{U_a}{U_a + U_b +U(Y)}\right],
\]
which means that $L_a$ dominates $L_b$ in RPS(1)-preferences.
\subsection{Risk tolerance}
In this section we show that if VNM utility is a linear function $U(x)=x$ and $L_a$ is a lottery with expected utility $U_a=\mathbb{E}[U(X)],$ then RPS(1) certainty equivalent $c$
verifies
\[
c< U_a, 
\]
if the lottery offers only positive payoffs, and 
\[
c> U_a,
\]
if the lottery offers only negative payoffs.
The certainty equivalent $c$ is defined by the equation
\[
\mathbb{E}\left[ \frac{e^c}{e^c+e^{U_a +X}}\right]{}={}\frac{e^{U_a}}{e^{U_a} + e^{2c}}.
\]
Here we using function $\Phi(x)=e^x$ that determines the stimuli increments in \eqref{RPS-1}. Consider first the gains: $X\geq0.$ Let $p(c)$ denote the left-hand side of the equation, and $q(c)$ the right-hand side. We have: $p'(c)>0,$ and  $q'(c)<0.$  Moreover,
\[
p(U_a){}={}\mathbb{E}\left[ \frac{1}{1+e^{X}}\right]\geq \frac{1}{1+e^{U_a}}=q(U_a),
\]
because function $(1+e^x)^{-1}$ is convex for $x\geq0.$ Thus, the point of intersection, $c,$ of graphs of $p$ and $q$ is less than $U_a.$ 

The other inequality is proved analogously, using the fact that $(1+e^x)^{-1}$ is concave for $x\leq 0.$

\subsection{RPS(1) preferences are not transitive}

Suppose, for simplicity, that VNM utility $U(c)=c.$ 
Consider two lotteries $L_a,$ $L_b,$ given by random payoffs $X$ and $Y$  that have the same expected utilities $U_a,$ $U_b,$ and have the same certainty equivalent in RPS(1) preferences. That is, 
\[
\mathbb{E}[X]=\mathbb{E}[Y]=U_a,
\]
and  there a \$c, such that  
\[
\mathbb{E}\left[ \frac{c}{U_a + c+ X} \right]{}={}
\mathbb{E}\left[ \frac{c}{U_a + c+ Y} \right].
\]
However, if $c\not=U_a,$ lotteries $L_a,$ $L_b$ are not necessarily equivalent, as the value of integral 
\[
\mathbb{E}\left[ \frac{U_a}{2U_a + X}\right]
\]
can be less, or greater, than that with random payoff $Y.$

\end{section}

\end{document}